\documentclass[doublecol]{epl2} 
\usepackage{amsmath,amssymb}

\DeclareTextSymbol{\degree}{OT1}{23}

\title{Novel zone formation due to interplay between sedimentation and phase ordering}
\shorttitle{Floating crystals} %Insert here a short version of the title if it exceeds 70 characters

\author{Mathieu Leocmach\inst{1} \and C. Patrick Royall\inst{2} \and Hajime Tanaka\inst{1}}
\shortauthor{M. Leocmach \etal}

\institute{                    
  \inst{1} Institute of Industrial Science, University of Tokyo - Meguro-ku, Tokyo 153-8505, Japan\\
  \inst{2} School of Chemistry, University of Bristol - Bristol BS8 1TS, UK
}
\pacs{82.70.Dd}{Colloids}
\pacs{47.57.ef}{Sedimentation and migration}
\pacs{64.60.Cn}{Order-disorder transformations}

\abstract{
Usually, we expect large particles to sediment faster than small of the same material. Contrary to this intuition, we report a dynamical competition between sedimentation and phase ordering which leads to smaller particles settling faster than larger ones. We access this phenomenon using suspensions of polymers and two colloidal species which we image with confocal microscopy. Polymers mediate attractions between colloids, leading to phase separation and crystallisation. We find that the dynamical interplay between sedimentation, phase separation and crystal nucleation underlie this phenomenon. Furthermore, under certain conditions we find a kinetic pathway leading to an apparent coexistence between a one component crystal and binary fluid of equal buoyancy. These findings may be relevant to the basic understanding of sedimentation-induced zone formation in nature and industrial applications.
}
\begin{document}

\maketitle

\section{Introduction}
Under a gravitational field, solid particles of hard (metals, semiconductors and oxides) or soft matter (colloids, macromolecules and biological materials) are inhomogeneously distributed along the vertical direction: This process is called sedimentation. Key early advances include the discovery of sedimentation-diffusion equilibrium by Jean Perrin one century ago \cite{perrin}, where the number density $\rho$ as a function of height $z$ is given by \cite{russel} $P(z)/k_{\rm B}T=\frac{1}{l_{\rm g}}\int_{z}^{h}\rho(z)dz$, where $P(z)$ is the (osmotic) pressure, $k_{\rm B}$ is Boltzmann's constant and $T$ is temperature. Sedimentation may be characterised by the gravitational length $l_{\rm g}=k_{\rm B}T/mg$, where $m$ is the buoyant mass of the colloidal particle and $g$ is the gravitational acceleration. $l_{\rm g}$ may be recast as a Peclet number ($Pe=\tau_{\rm D}/\tau_{\rm S}$), the ratio between the time $\tau_{\rm D}$ it takes a particle to diffuse its own radius and the time $\tau_{\rm S}$ to sediment its own radius. The Peclet number characterises the colloidal ($Pe\lesssim1$) and granular limit ($Pe\gg1$). An isolated spherical particle of diameter $\sigma$ settles under gravity, at an average (Stokes') velocity $u_0=\frac{mg}{3\pi\eta\sigma}$. Here $m=\delta\rho\pi\sigma^{3}/6$ with $\delta\rho$ the mass density difference between colloids and solvent and $\eta$ the viscosity of the solvent. This leads to $u_0\sim\sigma^{2}$: Larger particles sediment faster.

From this one-particle picture, we expect that a system with two or more species with different Stokes' velocity may separate during the process of sedimentation. However, reality is not necessarily that simple. In a suspension of a finite concentration, the kinetics of sedimentation are strongly influenced by hydrodynamic interactions between particles \cite{russel,segre2001}, and direct attractive interactions between the sedimenting particles can have a drastic effect and even qualitatively alter the simple picture above \cite{monchoJorda2009}. For example, in a colloid-polymer mixture (C+P), complex couplings between sedimentation and the polymer concentration come into play \cite{poon1999cpm}. The polymer chemical potential drives the depletion attraction between the colloids, yet the polymer free volume is itself coupled to a state point of the system \cite{lekkerkerker1992}. Reflecting this feature, predictions from free volume theory suggest that highly non-intuitive behaviour occurs even in equilibrium in the form of a colloidal liquid floating above a colloidal gas \cite{schmidt2004a}. Furthermore, it was shown that a system of C+P undergoes flocculation and then each flock settles much faster than individual particles, and can couple to phase ordering such as crystallisation during sedimentation \cite{dehoog2001}. In the sticky-sphere limit (where the size of the polymer is very small compared to the size of the colloid) flocculation can lead to dense colloidal crystals or kinetically arrested gels\cite{buzzaccaro:098301}.

\begin{figure}
\onefigure[width=8.5cm]{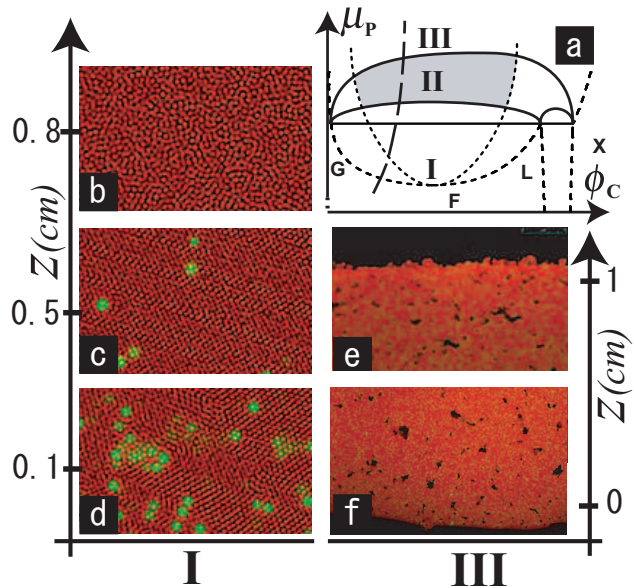} 
\caption{(colour online) (a) Schematic phase diagram of C+C+P showing the three regimes considered here, \textbf{I}, \textbf{II} and \textbf{III}. Long dashed line represents the values accessed by changing volume only. Systems initially in the shaded regions \textbf{II} eventually reach triple coexistence \cite{poon1999cpm}. (b-f) Final states without floating crystals, corresponding to regimes \textbf{I} ($\phi_{\rm P}^0=0.38$) (b,c,d) and \textbf{III} ($\phi_{\rm P}^0=0.73$) (e,f) in (a). Here (b) is fluid, and (c,d) are crystalline. Large particles are embedded in the crystal, a higher density towards the bottom (d) and reducing rapidly to a much lower density (c), reminiscent of a `sedimentation front'. (e,f) are sedimented gel with limited local crystallisation and trapped polymer-rich regions (black). Approximate heights of images are indicated on the vertical axes. Gravity points downwards.}
\label{fig:final states}
\end{figure}

Here we consider a system of binary colloid+polymer mixtures (C+C+P) with a very small overall concentration of one of the colloidal species. While the sedimentation behaviour of colloid-polymer mixtures has received some attention, its effect on binary hard-spheres is little understood. In bulk equilibrium binary hard spheres have been studied via experiment, \cite{bartlett1992,velikov2002,hobbie1998}, computer simulation \cite{eldridge1993edf} and theory \cite{lebowitz1964,mansoori1971,hansengoos2006,bartlett1990b}, and exhibit a plethora of plastic crystals \cite{williams2001}, binary crystal superlattices \cite{eldridge1993edf,bartlett1990b}, and Laves phases \cite{hynninen2007}. Here the interplay between sedimentation and the equilibrium phase diagram (see fig. \ref{fig:final states}(a) \cite{poon1999cpm}) and phase ordering leads to fascinating unexpected kinetic pathways. We emphasise that C+C+P is not merely a complication of C+P, but a minimal system exhibiting both phase condensation, driven by depletion, and frustration against crystallisation due to polydispersity. The competition between these two effects is crucial to understand the formation of sediment layers in nature. Moreover, we will show that it is in fact possible to gain much insight into the physics of C+C+P mixtures where small colloids dominate large ones, by considering the unperturbed C+P case for the initial kinetics and the binary hard sphere case to describe the dense sediment.

\section{Experimental}
We used poly(methyl methacrylate) (PMMA) colloids, sterically stabilised with polyhydroxyl steric acid. The large (L) and small colloids (S) have diameters of $\sigma_{\rm L}=1.20\:\mu$m and $\sigma_{\rm S}=0.680\:\mu$m ($\xi_{\rm SL}\equiv\sigma_{\rm S}/\sigma_{\rm L}=0.57$) and polydispersity 5\% and are labelled with 4-chloro-7-nitrobenzo-2-oxa-1,3-diazol (NBD) and rhodamine isothiocyanate, respectively. The solvent was cis-decalin whose relative dielectric constant is 2.4. Results on the electrostatic charging in solvents of comparable dielectric constants \cite{roberts2007} suggest that we expect only a handful ($<10$) charges per particle in this system, implying electrostatic interactions between colloids which are inferior to $k_{\rm B}T$.

The polymer used was polystyrene (PS) with a molecular weight of $3.1\times10^{7}$ ($M_{\rm w}/M_{\rm n}=1.3$). We estimate the polymer radius of gyration in an ideal solvent as $R_{\rm G}=160$ nm: the theta temperature of a PS/cis-decalin mixture is 12\degree{}C; following \cite{PhysRevE.52.5205} we assume that our system is reasonably well-described assuming polymer ideality. Some deviations are observed, particularly at higher colloid volume fraction \cite{aarts2002, royall2007c}, however, we argue that a scaled particle approach is appropriate for the level of this work. This assumption was discussed in more detail in \cite{royall2007c}. The polymer-large colloid size ratio is estimated as $\xi_{\rm PL}\equiv\sigma_{\rm P}/\sigma_{\rm L}=0.27$ implying a polymer-small colloid size ratio of $\xi_{\rm PS}=0.47$. According to \cite{lekkerkerker1992}, a S+P system would undergo gas-liquid phase separation, but not a L+P system, however experiments showed gas-liquid phase separation for a size ratio around $0.25$ \cite{Ilett1995}. Neither can be considered as the sticky sphere limit.

Gravitational lengths for large and small colloids are $l_{\rm g}^{\rm L}=1.75\:\mu$m and $l_{\rm g}^{\rm S}=9.64\:\mu$m respectively. Peclet numbers are $0.684$ and $0.0705$ respectively. We set the respective quantities of large and small particles in order to have $\phi_{\rm S}^0\ll\phi_{\rm L}^0$ ($\phi_{\rm S}^0=50\phi_{\rm L}^0$). Here the superscript 0 denotes the volume fraction at time=0, i.e., the mean colloid volume fraction. Similarly, to insure gas-liquid separation, we set the polymer concentration to keep $\phi_{\rm P}^0=4R_{\rm G}^{3}\rho_{\rm P}/3=1.8\phi_{\rm S}^0$, where $\rho_{\rm P}$ is the polymer number density. The solvent volume fraction is our experimental variable. Here we specify this solvent volume fraction by the polymer concentration $\phi_{\rm P}^0$. 

We used a Leica SP5 confocal microscope, using 488 nm and 532 nm laser excitation for the NBD and rhodamine labelled colloids respectively. Local volume fractions of large colloids were estimated by particle tracking. For the small colloids we used intensity measurements which were calibrated against homogeneous samples of known concentrations. In our system, we found a linear dependence of measured intensity against colloid volume fraction in the fluid but not in the crystal. The volume occupied locally by the crystals was given by image segmentation. Rectangular sample cells (height: 40 mm, width: 5.0 mm and depth 2.0 mm) were used. Sedimentation is a slow process and we use the term ``final state'' to express an apparent steady state after the time needed for a single particle to sediment from the middle to the bottom of the cell (a few weeks). There is practically no observable change after two weeks. This final state is \emph{not} the lowest free energy state, the time needed to reach the free energy global minimum runs to years, due to the slow diffusion of the particles in the dense sediment.

\section{Results}
Since the vast majority of the colloids are small, it is instructive to consider the S+P system [fig. \ref{fig:final states}(a)], whose kinetic (and equilibrium) behaviour has been studied \cite{poon1999cpm}. We identify three regimes of initial composition, \textbf{I} leading to colloidal fluid or phase separation into a colloid-rich and colloid-poor fluid phases, \textbf{II} leading to triple coexistence \cite{poon1999cpm} and \textbf{III} leading to arrested phase separation (gel). We distinguish regimes \textbf{I} and \textbf{II} by whether the system undergoes homogeneous nucleation or not. In fact, regime \textbf{I} has the potential for colloidal gas-liquid type phase separation to dominate the early kinetics, which is similar to regime \textbf{II}. However, as we discuss below, it is in the longer-time kinetics that a clear difference between regimes \textbf{I} and \textbf{II} is seen.

In regime \textbf{I}, the final state resembles a binary hard sphere fluid (BHS): fluid at the top with a small $\phi_{\rm L}$ and crystals of S particles (X$_{\rm S}$) at the bottom with very few grain boundaries and some large particles included individually as defects in the X$_{\rm S}$ lattice. These defects were more concentrated at the bottom of the crystals, as expected. Due to the small $\phi_{\rm L}^0$, no X$_{\rm L}$ is observed. In other words, without phase separation of the majority (small) species, the behaviour is consistent with the naive expectation above [fig. \ref{fig:final states}(b,c,d)]. Experimentally, by changing $\phi_{\rm P}^0$ from 0.40 to 0.43, we cross the boundary between regime \textbf{I} and \textbf{II}. Note that in the underlying S+P phase diagram \cite{poon1999cpm}, the polymer concentration of $\phi_{\rm P}^0\sim0.40$ is required for triple coexistence. In regime \textbf{III,} with a high polymer concentration ($\phi_{\rm P}^0\geq0.73$), we observed a gel of S with some evidence of local crystallinity coexisting with a colloidal gas, [fig. \ref{fig:final states}(e,f)]. 

\begin{figure}
\onefigure[width=7.cm]{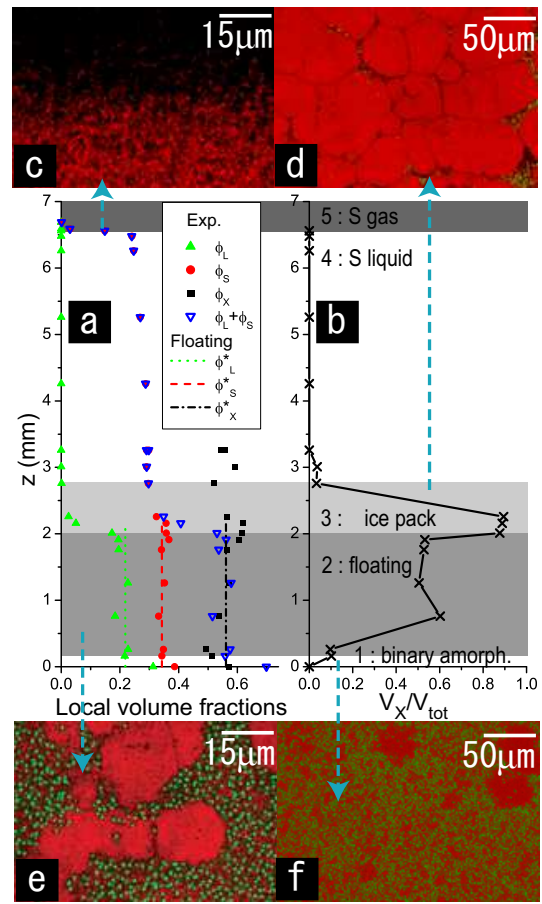} 
\caption{(colour online) (a) Sedimentation profile of a C+C+P sample (regime \textbf{II}) during sedimentation, with its bottom part already in the final state. Symbols represent the measured local volume fractions of each component in each phase. Lines are the theoretical floating volume fractions. (b) The fraction of the volume occupied by X$_{\rm S}$. Overall composition:$\phi_{\rm L}^0=0.004$, $\phi_{\rm S}^0=0.186$ and $\phi_{\rm P}^0=0.45$. Total size of the sample: 3 cm (only the bottom is shown here). (c) gas (zone 5)-liquid (zone4) interface. (d) ice pack (zone 3). (e) floating crystals (zone 2). (f) glass with a few crystallites (zone 1). Bright (red) regions are X$_{\rm S}$. Very bright (green) dots are large particles.}
\label{fig:xp-sedim-profile}
\end{figure}

\section{Regime II}
More intriguing is the final state of C+C+P between the two previous cases, regime \textbf{II}, to which we devote the remainder of the discussion. A simple equilibrium expectation would be a final state very similar to regime I, but with a gas-liquid interface. However, we observed the following multi-zone structure from bottom to top (see fig. \ref{fig:xp-sedim-profile}). At the bottom of the sediment, we find a binary amorphous solid (zone 1) [fig. \ref{fig:xp-sedim-profile}(f)], followed by a coexistence between X$_{\rm S}$ and a binary fluid (zone 2) [fig. \ref{fig:xp-sedim-profile}(e)]. Crystals of small particles are limited in extent (20$\sim$30 $\mu$m) and the binary fluid between them percolates from the upper to the lower bound of this zone. Neither melting nor growth of the crystals was detected, suggesting that crystal and fluid are really coexisting `thermodynamically'. This is further supported by Supplementary Movie 1 which reveals diffusion in the binary fluid, showing that the system has some degree of thermal motion and is not completely jammed. (zone 3) Crystals of small particles almost fill space but with a lot of grain boundaries {[fig. \ref{fig:xp-sedim-profile}(d)]. The size of each single crystal is 20-100 $\mu$m. $\phi_{\rm L}$ is very small. We refer to this zone as the ``ice pack'' because it is a crystalline state situated over a phase with a lot of fluid. (zone 4) The liquid phase is dominated by small colloids. At high altitude, we recover the gas phase of small colloids (zone 5) [fig. \ref{fig:xp-sedim-profile}(c)].

The composition profile and the volume occupied by the crystals are shown as a function of height for each phase in figs. \ref{fig:xp-sedim-profile}(a) and (b), respectively. The present data were taken before the complete sedimentation of zone 4, but samples observed 3 months later showed no quantitative changes in their lower part (zones 1-3). Due to further crystallisation, the upper bound of the ``ice pack'' was higher and the liquid extended only $\sim10$ gravitational length before the gas-liquid interface. This is consistent with simple C+P mixtures \cite{poon1999cpm}.
 
The most striking feature of the multi-zone structure is that the compositions of the coexisting phases ($\phi_{\rm L}=0.22\pm0.02$, $\phi_{\rm S}=0.34\pm0.02$, $\phi_{\rm X}=0.56\pm0.02$) are almost constant over zone 2. We note that the \emph{macroscopic} vertical extent of zone 2, ($\sim2$ mm), is thousands times of the colloidal gravitational length. This indicates that the two coexisting phases, X$_{\rm S}$ and the binary fluid have the same density, otherwise they should be separated under the gravitational field. This is confirmed graphically on fig. \ref{fig:xp-sedim-profile}(a) by the correspondence between $\phi_{\rm L}+\phi_{\rm S}$ and $\phi_{\rm X}$.

\begin{figure*}[ht]
\onefigure[width=13cm]{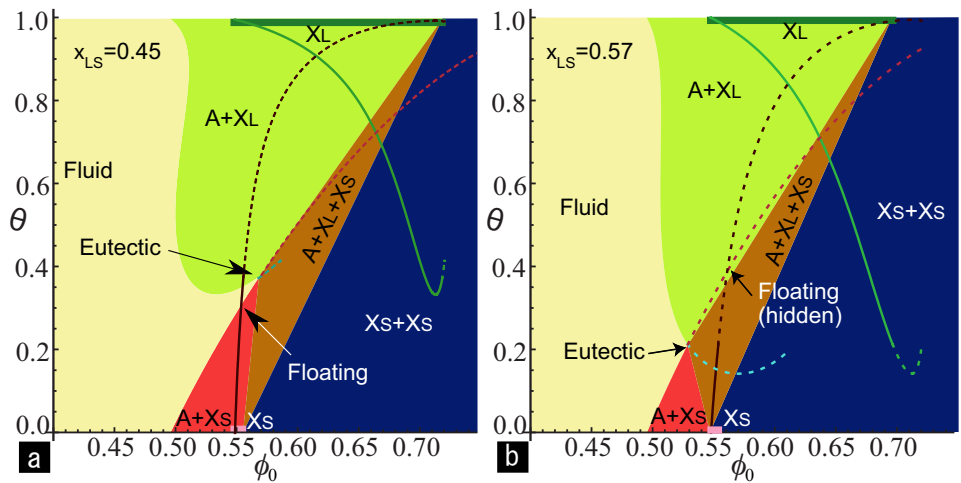}
\caption{(colour online) Theoretical phase diagram of binary hard spheres for $\xi_{\rm SL}=0.45$ and $\xi_{\rm SL}=0.57$. Melting curves are represented by continuous lines as a function of $\theta=\phi_{\rm L}/\phi_0$ \emph{in the coexisting fluid}. Dashed curves are prolonging phase boundaries and melting curves unstable to further phase separation.}
\label{fig:PhaseDiagramBHS}
\end{figure*}

\section{Phase behaviour}
Now we consider how such floating crystals may be rationalised, taking a binary hard sphere mixture (BHS) as a minimal system exhibiting this phenomenon. The more involved C+C+P, whose essential features are similar for these purposes to BHS will be presented elsewhere. However, we note that the inclusion of polymers tends to destabilise binary crystals, which we neglect here. Furthermore, no binary crystals were observed in the experiments. Following \cite{bartlett1990b}, in fig. \ref{fig:PhaseDiagramBHS} we focus on the binary fluid (index A), the two possible one component crystals (index X) and their coexistences, including the three-phase eutectic region.

Let us consider a monodisperse fluid-crystal coexistence of hard spheres of diameter $\sigma_{\rm S}=1$. The buoyancy difference between the two coexisting phases is proportional to their volume fraction difference $\Delta\phi=\phi_{\rm X}-\phi_{\rm A}>0$. If we add one sphere of diameter $\sigma_{\rm L}=2\sigma_{\rm S}$ in the fluid and keep the osmotic pressure balance against the same crystal, we have to remove one small sphere to keep the same number density. But the large particle occupies more volume than the small particle that was removed so the volume fraction in the fluid increases. So $\Delta\phi$ decreases with the introduction of large particles in the fluid. This effect should lead ultimately to $\Delta\phi=0$ for a given $\phi_{\rm L}^\ast$. If both colloidal species are the same material, both phases have the same buoyancy, just as a zeppelin floats in the air. This is a single point on the phase diagram because $\phi_S^*$ and $\phi_{\rm X}^\ast$ are fixed due to coexistence relations: the ``floating point''. For a crystal of large particles coexisting with a binary fluid, ${\rm A}_{\rm L+S}\leftrightarrow {\rm X}_{\rm L}$, adding smaller particles to the fluid enhances $\Delta\phi$, consistent with \cite{dijkstra1999}.

More quantitatively, it is possible to integrate a binary hard sphere fluid equation of state (here Hansen-Goos-Roth \cite{hansengoos2006}) and a hard sphere crystal equation of state (here Alder-Wainwright \cite{alder1957}) to obtain the reduced excess chemical potentials in both phases\cite{bartlett1990b}. For the fluid we get
\begin{align}
&\mu_{\rm A,L}^{ex}(\phi_{\rm L},\phi_{\rm S}) =\notag\\
&\frac{\phi_3}{1-\phi_0}+\frac{5-\phi_3}{1-\phi_0}(\phi_1+\phi_2)-\frac{2(1-3\phi_0)}{\phi_0(1-\phi_0)^2}\phi_1\phi_2 \notag\\
&+ 3\frac{1-3\phi_0+\phi_0^2}{\phi_0(1-\phi_0)^2}\phi_1^2+\frac{2-5\phi_0+6\phi_0^2-\phi_0^3}{\phi_0^2(1-\phi_0)^3}\phi_1^3 \notag\\
&- \frac{(\phi_0-\phi_1)(\phi_0^2+2\phi_1^2-\phi_0(\phi_1-2\phi_2))}{(1-\phi_0)^3}\ln(1-\phi_0), \label{eq:muexAL}
\end{align}
\begin{align}
&\mu_{\rm A,S}^{ex}(\phi_{\rm L},\phi_{\rm S}) =\notag\\
&\frac{\xi_{\rm SL}^3\phi_3}{1-\phi_0}+\frac{5-\phi_3}{1-\phi_0}\xi_{\rm SL}(\phi_1+\xi_{\rm SL}\phi_2)-\frac{2(1-3\phi_0)}{\phi_0(1-\phi_0)^2}\xi_{\rm SL}^3\phi_1\phi_2 \notag\\
&+ 3\frac{1-3\phi_0+\phi_0^2}{\phi_0(1-\phi_0)^2}\xi_{\rm SL}^2\phi_1^2+\frac{2-5\phi_0+6\phi_0^2-\phi_0^3}{\phi_0^2(1-\phi_0)^3}\xi_{\rm SL}^3\phi_1^3\notag\\
&- \frac{(\phi_0-\xi_{\rm LS}\phi_1)(\phi_0^2+2\xi_{\rm SL}^2\phi_1^2-\phi_0(\phi_1-2\xi_{\rm LS}\phi_2))}{(1-\phi_0)^3}\notag\\
&\qquad\times\ln(1-\phi_0). \label{eq:muexAS}
\end{align}
with $\phi_i=\phi_{\rm L}+\frac{\phi_{\rm S}}{\xi_{\rm SL}^i}$, $i=0...3$, the weighted volume fractions. Unlike \cite{schmidt2004a} we disregard gravity because we are looking for two equally dense phases. At ${\rm A}_{\rm L+S}\leftrightarrow {\rm X}_{\rm L}$ coexistence, we have equality of the pressure and of the chemical potential of L in each phase:
\begin{align}
\mu_{\rm A,L}^{\rm ex}(\phi_{\rm L},\phi_{\rm S}) &= \mu_{\rm X_{\rm L}}^{\rm ex}(\phi_{\rm X_{\rm L}})+\ln\frac{\phi_{\rm X_{\rm L}}}{\phi_{\rm L}}\label{eq:equil1},\\
p_{\rm A}(\phi_{\rm L},\phi_{\rm S}) &= p_{\rm X_{\rm L}}(\phi_{X_{\rm L}}). \label{eq:equil2}
\end{align}

Starting from the known monodisperse coexistence, we integrate eqs. (\ref{eq:equil1}) and (\ref{eq:equil2}) numerically using eq. (\ref{eq:muexAL}). This gives us the parametric representation of the ${\rm X}_{\rm L}$ crystallisation line in the $(\phi_{\rm L},\phi_{\rm S})$ plane. The X$_{\rm S}$ crystallisation line is similarly obtained. Being interested in the respective buoyancy of the phases, we plot the phase diagram (fig. \ref{fig:PhaseDiagramBHS}) in the $(\phi_0,\theta)$ plane, with $\phi_0=\phi_{\rm L}+\phi_{\rm S}$ the total volume fraction and $\theta=\phi_{\rm L}/\phi_0$ the volume proportion of L particles. Of course, $\theta=1$ in the crystal X$_{\rm L}$ and 0 in X$_{\rm S}$. In order to display the coexisting phases in fig. \ref{fig:PhaseDiagramBHS}, the melting lines of the two crystals were drawn as follows: the crystal compositions $\phi_{\rm X_{\rm L}}$ and $\phi_{\rm X_{\rm S}}$ are drawn with respect to $\theta$ \emph{in the coexisting fluid}. Therefore, on the abscissa we see the respective buoyancies of the fluid and coexisting crystal. The crossing of the two curves indicates the floating point.

As expected, $\Delta\phi$ is broadened by adding S in the ${\rm A}_{\rm L+S}\leftrightarrow {\rm X}_{\rm L}$ coexistence. In the ${\rm A}_{\rm L+S}\leftrightarrow {\rm X}_{\rm S}$ coexistence, adding L shrinks the buoyancy gap, leading to a floating point whose position depends on the size ratio $\xi_{\rm SL}$ (see fig. \ref{fig:evolutionQ}). Moreover, for $\xi_{\rm SL}>\xi_{\rm SL}^{\rm max}\simeq0.464$, the floating point is not located in the stable ${\rm A}_{\rm L+S}\leftrightarrow {\rm X}_{\rm S}$ coexistence region, but unstable against further phase separation (${\rm X}_{\rm L}$ crystallisation). This is the case ($\xi_{\rm SL}=0.57$) our experimental system belongs to [see fig. \ref{fig:PhaseDiagramBHS} (b)]. Nevertheless, the floating point was realised in our experiments, suggesting the agreement between experimental and theoretical floating compositions shown in fig. \ref{fig:xp-sedim-profile} (a). No ${\rm X}_{\rm L}$ crystallisation was detected in the experimental time window.

\begin{figure}[h]
\onefigure[width=6.5cm]{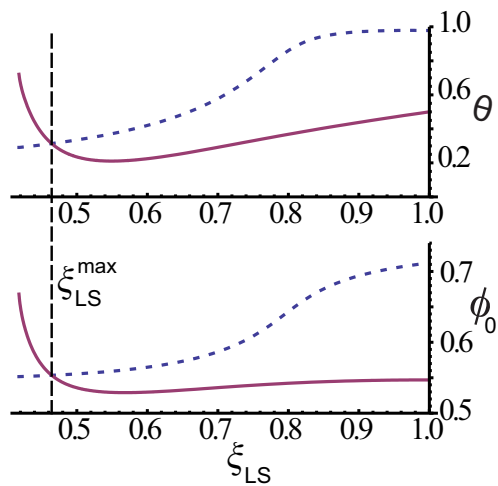}
\caption{$\xi_{\rm SL}$ dependence of eutectic (line) and floating (dash) proportion of L in volume (top) and total volume fraction (bottom). Inversion appends for $\xi_{\rm SL}=\xi_{\rm SL}^{\rm max}\simeq0.464$.}
\label{fig:evolutionQ}
\end{figure}

\section{Kinetics}
We now describe the dynamic sequence by which a sample in regime \textbf{II} reaches the final multi-zone structure. The initially homogeneous sample first undergoes a gas-liquid phase separation, induced by the depletion attraction. Then, droplets of colloidal liquid sediment. After 2 days, the sample is separated into a colloidal gas in the upper part of the sample and binary liquid in the lower part. We did not see any heterogeneous nucleation of crystals at the bottom, unlike regime \textbf{I}. We believe this is because the bi-dispersity dramatically suppresses nucleation of crystals in the bottom binary fluid. Without flocculation, particles in this fluid should sediment individually. After another 2 days, the higher sedimentation speed of the large particles leads to a zone at the top of the liquid, almost devoid of large particles. The sample is thus divided into three zones: colloidal gas at the top, liquid with few large particles in the middle, and a binary fluid rich in large particles at the bottom.

Above the sedimentation front of large particles, $\phi_{\rm L}$ is low so homogeneous nucleation of S crystals proceeds. Here we emphasize that this removal of large particles from a liquid phase of small particles due to sedimentation is crucial for the initiation of homogeneous nucleation of S crystals and the resulting inverted distribution of large and small particles. Reduction of frustration against crystallization is a necessary condition for frequent crystal nucleation to be realised \cite{Tanaka1999top,Royall2008}. Considering the buoyancy difference between a crystal nucleus of diameter $\alpha\sigma_{\rm S}$ and of volume fraction $\phi_{\rm X}$ and the surrounding S fluid ($\phi_{\rm S})$, the Peclet number of the nucleus can be expressed function of the Peclet number of a single particle S: $Pe_{\rm X}=\alpha^{4}(\phi_{\rm X}-\phi_{\rm S})Pe_{\rm S}$. For $\phi_{\rm X}-\phi_{\rm S}=0.1$, a crystal nucleus of a diameter larger than $3.4\,\sigma_{\rm S}$ has $Pe_{\rm X}>1$. Thus, such crystals experience little Brownian motion and simply sediment quickly. However, when these crystals reach the sedimentation front of the large particles [fig. \ref{fig:Lfront+diving}(a)], the sedimentation velocity drastically reduces, as the binary fluid has a higher total volume fraction and is therefore denser than the almost monodisperse fluid. The relevant Peclet number for the crystals is then $Pe_{\rm X}=10^{4}(\phi_{\rm X}-\phi_0)Pe_{\rm S}$. If the fluid and crystal compositions are close to the floating point, $\phi_{\rm X}\approx\phi_0$, which means $Pe_{\rm X} \ll 1$, the crystals float. After another 3 days, the crystals piled up from the bottom to the top of the large sphere sediment [fig. \ref{fig:Lfront+diving}(b)]. The crystals did not grow after entering the binary fluid zone and we conclude that X$_{\rm S}$ and the binary fluid appear to ``coexist''.

When the pile of the floating crystals reaches the top of the large-sphere-rich sediment, crystal nucleation is still active in the upper monodisperse $S$ fluid. So crystals continue to pile up above the large sphere sediment. In the monodisperse fluid, however, there is no mechanism which prevents further growth of the crystals, and thus they fill all the available space to form the ``ice pack'' [see fig. \ref{fig:xp-sedim-profile}(c)]. This fluid-crystal phase separation finally stops when the monodisperse liquid layer reaches a stable depth of a few gravitational lengths, which leads to the final state.

\begin{figure}
\onefigure[width=8.5cm]{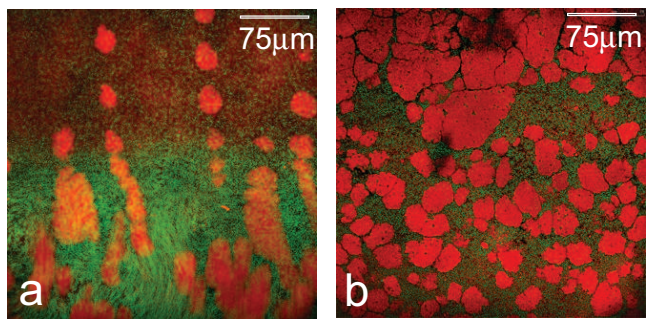}
\caption{(colour online) The kinetics of phase ordering and sedimentation in a phase separating C+C+P system, regime \textbf{II}: ($\phi_{\rm P}^0 = 0.36$) (a) Sedimentation front of large particles in the fluid. Time-lapse (every 10 min) picture showing crystals of small particles falling into the binary fluid. Total time = 3 days, total image width 0.1 mm. See also Supplementary Movie 2. (b) The same place in the final state (one month), covered by the ice pack.}
\label{fig:Lfront+diving}
\end{figure}

\section{Conclusions}
To sum, we found that the introduction of attractive interactions between binary colloid mixtures leads to complex dynamic interplay between sedimentation and phase ordering. This kinetic pathway is fundamentally different from that of sedimenting binary hard sphere mixtures, where sedimentation is followed by heterogeneous nucleation: crystallisation is followed by sedimentation. Despite the complex nature of this multicomponent C+C+P system, we are able to rationalise many of our findings by appeal to previous C+P work, due to the low overall concentration of the larger colloids, and to binary hard spheres, due to the low polymer concentration in the colloid-rich phase after phase separation. The dynamic sequence we have observed leads to the formation of a multi-zone structure, including an exotic floating crystal zone and an ice pack zone. This final multi-zone state is a metastable state, likely connected to a hidden floating point. This metastability of the floating crystal zone is associated with the inability of the small particle crystallites to fully coarsen into a continuous and homogeneous phase and thus an entirely kinetic effect. After a long time, we expect crystals in the ice pack zone to grow at the expense of small crystallites in the floating zone to reduce the interfacial energy. This suggests that the subtle interplay between metastability and kinetic arrest, which is a signature trait of soft matter that can also be found all around us in the real world, can play a significant role in dynamical phenomena including sedimentation. Our study indicates that the introduction of phase ordering in sedimentation leads to a rich variety of the final zone formation, which may contribute to the deeper understanding of zone formation industrial applications and the natural world.

\acknowledgments
We thank Jeroen van Duineveldt for helpful discussions. ML thanks MEXT for a scholarship. CPR acknowledges the Royal Society for financial support. HT acknowledges a grant-in-aid from MEXT.

\end{document}